\documentclass[aps,prl,twocolumn,superscriptaddress,showpacs]{revtex4-1}
 
\usepackage{graphics}
\usepackage{graphicx}  
\usepackage{color}
\usepackage[english]{babel}

\ifx\pdftexversion\undefined
\usepackage[dvips]{hyperref}
\else
\usepackage{hyperref}
\fi
\hypersetup{
  colorlinks = true, linkcolor = blue
}

\begin{document}

\title{Hanbury-Brown and Twiss bunching of phonons and of the quantum depletion in a strongly-interacting Bose gas}

\author{Hugo Cayla}
\affiliation{Universit\'e Paris-Saclay, Institut d'Optique Graduate School, CNRS, Laboratoire Charles Fabry, 91127, Palaiseau, France}
\author{Salvatore Butera}
\affiliation{School of Physics and Astronomy, University of Glasgow, Glasgow G12 8QQ, UK}
\author{C\'ecile Carcy}
\affiliation{Universit\'e Paris-Saclay, Institut d'Optique Graduate School, CNRS, Laboratoire Charles Fabry, 91127, Palaiseau, France}
\author{Antoine Tenart}
\affiliation{Universit\'e Paris-Saclay, Institut d'Optique Graduate School, CNRS, Laboratoire Charles Fabry, 91127, Palaiseau, France}
\author{Ga\'etan Herc\'e} 
\affiliation{Universit\'e Paris-Saclay, Institut d'Optique Graduate School, CNRS, Laboratoire Charles Fabry, 91127, Palaiseau, France}
\author{Marco Mancini}
\affiliation{Universit\'e Paris-Saclay, Institut d'Optique Graduate School, CNRS, Laboratoire Charles Fabry, 91127, Palaiseau, France}
\author{Alain Aspect}
\affiliation{Universit\'e Paris-Saclay, Institut d'Optique Graduate School, CNRS, Laboratoire Charles Fabry, 91127, Palaiseau, France}
\author{Iacopo Carusotto}
\affiliation{INO-CNR BEC Center and Dipartimento di Fisica, Universit\`a di Trento, I-38123 Povo, Italy}
\author{David Cl\'ement}
\affiliation{Universit\'e Paris-Saclay, Institut d'Optique Graduate School, CNRS, Laboratoire Charles Fabry, 91127, Palaiseau, France}
\affiliation{Institut Universitaire de France (IUF)}

\date{\today}

\begin{abstract}
We report the realisation of a Hanbury-Brown and Twiss (HBT)-like experiment with a gas of strongly interacting bosons at low temperatures. The regime of large interactions and low temperatures is reached in a three-dimensional optical lattice and atom-atom correlations are extracted from the detection of individual metastable Helium atoms after a long free-fall. We observe a HBT bunching in the non-condensed fraction of the gas whose properties strongly deviate from the HBT signals expected for non-interacting bosons. In addition, we show that the measured correlations reflect the peculiar quantum statistics of atoms belonging to the quantum depletion and of the Bogoliubov phonons, {\it i.e.}, of collective excitations of the many-body quantum state. Our results demonstrate that atom-atom correlations provide information about the quantum state of strongly-interacting particles, extending the interest of HBT-like experiments beyond the case of non-interacting particles. 
\end{abstract}

\maketitle

In systems of non-interacting and indistinguishable quantum particles, correlations are rooted in quantum statistics. A paradigmatic example is the  bunching of photons received from a source, as described by Glauber's quantum formalism \cite{glauber1963} to interpret the Hanbury-Brown and Twiss (HBT) observation of intensity  \cite{hanbury1956} or photon \cite{rebka1957} correlations. Such a method yields information both on the photon statistics (chaotic {\it vs.}  fully coherent \cite{arecchi1966}) and on the spatial distribution of the emitters, i.e., on the size of the source \cite{hanbury1956i}. 

This approach pioneered by Hanbury-Brown and Twiss was successfully extended to characterize quantum states in various situations, ranging from high-energy physics \cite{baym1998} and solid-state devices \cite{oliver1999, henny1999} to cold atoms \cite{ottl2005, schellekens2005, folling2005, rom2006, jeltes2007, dall2013, preiss2019}. In non-interacting atomic gases at thermal equilibrium, the bunching (for bosons) \cite{ottl2005, schellekens2005, folling2005, dall2013} and anti-bunching (for fermions) \cite{rom2006, jeltes2007, preiss2019} is set by the quantum statistics and the thermal occupation of single-particle states. On the other hand, atom-atom correlations are absent in a fully coherent Bose-Einstein Condensate (BEC) \cite{ottl2005, schellekens2005}, in analogy with the lack of photon-photon correlations in a single mode laser beam \cite{arecchi1966}.

In contrast, HBT-like measurements with interacting particles are scarce. In optics, the role of non-linearities during the propagation from the source to the detector was studied \cite{bromberg2010}, but interacting photon fluids as a source \cite{carusotto2013} have not yet been probed. With atoms, two-body correlations were used to characterize the coherence properties of weakly-interacting Bose gases across the Bose-Einstein transition \cite{perrin2012}, in a regime where the temperature exceeds the interaction energy. In the opposite regime where interactions dominate, one may observe the interplay between quantum statistics and interactions in many-body systems. This is the goal of the experiment presented here. 

In this letter, we report on the measurement of momentum-momentum correlations in an ensemble of interacting atoms, in the low-temperature regime dominated by interactions. This regime is achieved by using a three-dimensional (3D) optical lattice to enhance the interactions. The two-body correlations are extracted from detecting individual metastable Helium atoms ($^4$He$^*$) after a long free-fall \cite{cayla2018, tenart2020}. We characterize the bunching properties -- amplitude and width of the HBT bump --  and highlight the differences with previous findings in non-interacting ensembles. These differences are attributed to the properties of the quantum depletion and of the collective excitations  -- Bogoliubov phonons -- in the ensemble of interacting atoms probed in our experiment.

In analogy with HBT experiments with an incoherent source of light, where photon correlations are measured in far-field, HBT experiments with thermal He$^*$ gases \cite{schellekens2005, jeltes2007, dall2013} look for atom correlations after a long free-fall, {\it i.e.}, in the basis of single-particle momentum states $|{\bf k} \rangle$. For each sample released onto a position- and time-resolved detector sensitive to individual He$^*$ atoms, the atom distribution is recorded in 3D, and the two-body correlation function for that sample is calculated over a volume of interest $\Omega_\mathbf{k}$. Repeating the procedure with many samples, one averages to obtain an experimental evaluation of the volume integrated two-body correlation function \cite{naraschewski1999}
\begin{equation}
\label{eq:g2}
g_{\Omega_{\bf k}}^{(2)}( {\bf \delta k}) = \frac{\int_{\Omega_{\bf k}}  \langle a^{\dagger}({\bf k}) a^{\dagger}({\bf k}+{\bf \delta k}) a({\bf k}) a({\bf k}+{\bf \delta k}) \rangle \, \mathrm{d} {\bf k}}{\int_{{ \Omega_{\bf k}}} \langle n({\bf k}) \rangle \langle n({\bf k}+{\bf \delta k})\rangle \, \mathrm{d} {\bf k}}
\end{equation}
where $a^{\dagger}({\bf k})$ (resp. $a({\bf k}))$ is the creation (resp. annihilation) operator associated with momentum ${\bf k}$. 

In an ideal (non-interacting) and non-condensed Bose gas at thermal equilibrium, one expects $g^{(2)}(0)=2$ because of (chaotic) Gaussian statistics. The Gaussian nature of the statistics derives from the random and uncorrelated populations of the momentum states at thermal equilibrium \cite{aspect2016, carcy2019}. Moreover, the in-trap density of a non-condensed Bose gas has a Gaussian shape with a rms size $s_{\rm th}=\sqrt{k_{\rm B}T/m \omega^2}$, and the bunching bump has also a Gaussian shape with a width (half-width at 1/e)  $\sigma_k^{\rm ideal}=1/s_{\rm th}$ \cite{schellekens2005, aspect2016}. This relation is analogous to that used to deduce the angular size of a star from the intensity HBT correlation length \cite{hanbury1956i}. This description of atom correlations also applies to weakly-interacting bosons when the temperature largely exceeds the interaction energy  \cite{schellekens2005,dall2013}.
 
\begin{figure}[ht!]
\includegraphics[width=\columnwidth]{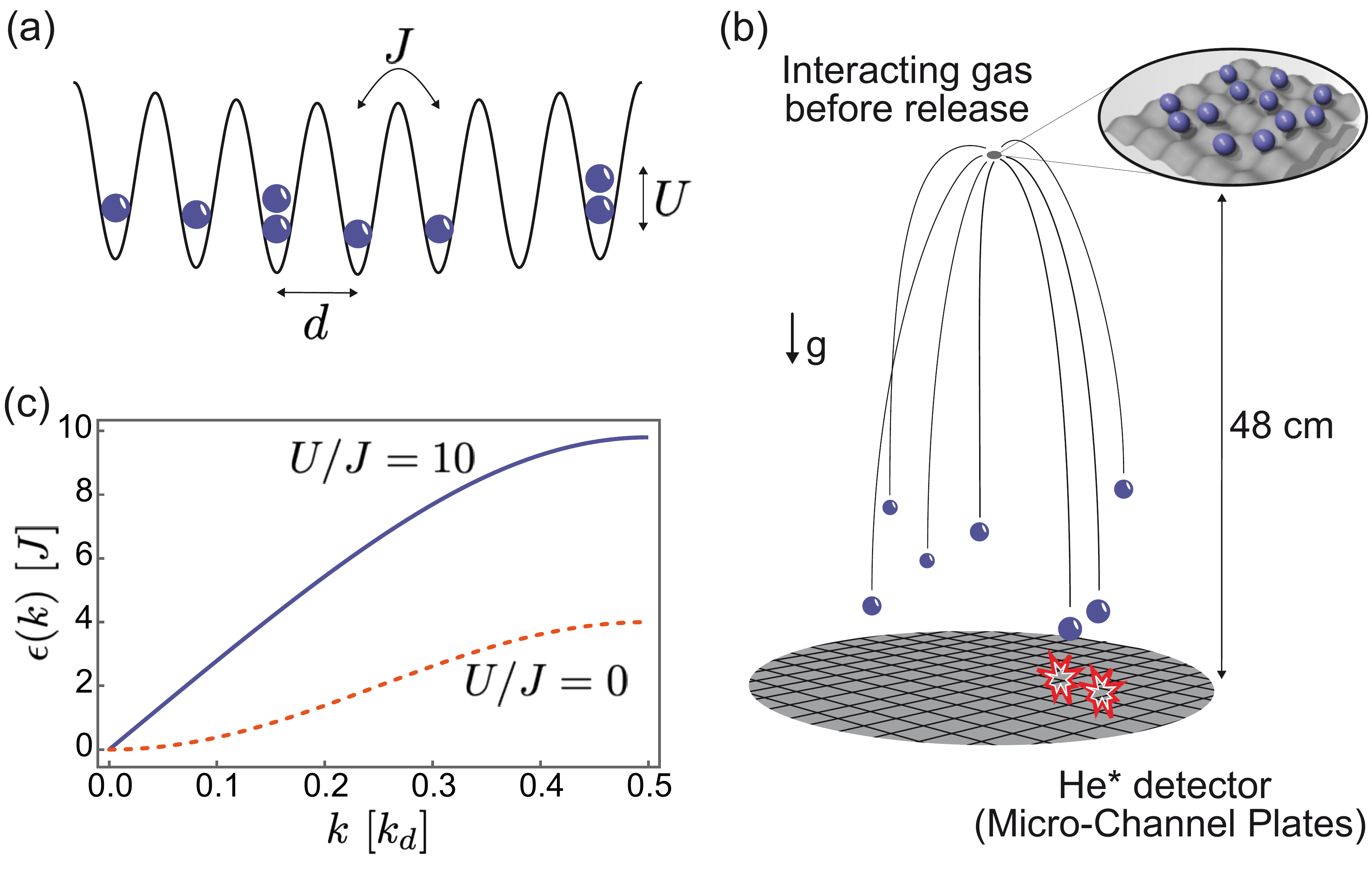}
\caption{(a) Interacting metastable Helium $^4$He$^*$ atoms are loaded into a 3D optical lattice of spacing $d=775~$nm. The tunnelling energy is denoted $J$ and the on-site interaction energy $U$. In this work, we set $U/J=10$. (b) Sketch of the detection method. Atoms are released from the lattice and reach the He$^*$ detector after a long free-fall (325 ms) that maps the momentum distribution of the trapped atoms into the measured spatial distribution. The use of a micro-channel plate, in combination with delay-line anodes (not shown), allows for the detection of individual atoms in three-dimensions \cite{cayla2018} from which the HBT correlations are extracted. (c) Dispersion relation of the Bogoliubov excitations of the interacting gas in the lattice (solid blue line, $U/J=10$) and of non-interacting particles in the lattice (dashed orange line, $U/J=0$). The collective nature of the excitations is dominant in the linear part of the Bogoliubov spectrum at low energies (up to $k\simeq0.35 k_{d}$ here) which is associated with phonons. 
}
\label{fig1}
\end{figure}

In the opposite low-temperature regime where interactions dominate, the bunching properties differ because interactions affect the population statistics of the $|{\bf k} \rangle$-states. To get some insight into this complex physics, we use the Bogoliubov approximation for interacting bosons \cite{bogoliubov1947}. The interacting Bose gas is described as a many-body ground-state -- the condensate and  the  quantum depletion -- and an ensemble of non-interacting (quasi-particle) excitations. For energies up to the interaction energy $\mu$ in the condensate, these Bogoliubov excitations have the collective nature of phonons. They are non-interacting bosons with populations set by the temperature. Their statistics is therefore that of ideal bosons, {\it i.e.} Gaussian. Because the Bogoliubov transform between particle and quasi-particle operators is linear, the statistics of the particle momenta is Gaussian as well, hence $g^{(2)}({\bf \delta k}={\bf 0})=2$ \cite{toth2008, mathey2009, butera2020}. Quite remarkably, a bunching at ${\bf \delta k}={\bf 0}$ is also expected for atoms belonging to the quantum depletion, although they form a pure state of pair-correlated atoms with opposite momenta. That bunching stems from the fact that when one observes atoms with momenta almost equal, the correlations are measured between atoms belonging to two different pairs. The density matrix describing these atoms, obtained by tracing over the second partners of each pair, which are ignored, has a chaotic character. This origin is analog to that of the thermal (chaotic) statistics encountered when one observes only one partner of parametric-down conversion photon pairs \cite{yurke1987} or of atom pairs produced in a two-body collision process \cite{molmer2008, perrier2019}.

For an interacting gas described within the Bogoliubov approximation, the width $\sigma_k^{\rm B}$ of the two-body correlation bump in $\mathbf{k}$-space is expected to behave differently from the $1/\sqrt{T}$ variation of an ideal thermal gas \cite{butera2020}. At $T=0$, the only states relevant to the bunching belong to the quantum depletion. Their spatial in-trap extent, which is limited to the BEC radius $R_{\rm bec}$, determines $\sigma_k^{\rm B}(T=0)$. Since the BEC rms size is $\sim R_{\rm bec}/\sqrt{2}$, a rough estimate is $\sigma_k^{\rm B}(T=0) \sim \sqrt{2}/R_{\rm bec}$, in analogy with $\sigma_k^{\rm ideal}=1/s_{\rm th}$. At small non-zero temperatures ($k_{B}T\ll\mu$), low-lying phononic excitations are populated, whose spatial in-trap size hardly extends beyond $R_{\rm bec}$ as well. When the temperature increases, the Bogoliubov excitations progressively extend out of the condensate and $\sigma_k^{\rm B}(T)$ should slowly decrease with increasing $T$. In the low temperature regime, both contributions associated with the atoms interaction are thus expected to give a bunching whose width is definitely smaller than that of an ideal thermal gas.

The above description for harmonically trapped interacting bosons also applies when a 3D shallow optical lattice is present. In shallow lattices, the ratio $U/J$ of the on-site interaction energy $U$ to the tunnelling amplitude $J$ (see Fig.~\ref{fig1}a) is smaller than that of the Mott transition, and the gas is Bose-condensed at low temperatures. Accounting for the lattice with the effective mass $m^*=\hbar^2/2 J d^2$ ($d$ is the lattice spacing, corresponding to a momentum $k_{d}=2\pi/d$) and trap frequency $\omega^*=\omega \sqrt{m/m^*}$ \cite{kramer2002}, the Bogoliubov description is unchanged. Moreover, the bunching properties of ideal lattice bosons are identical to those of ideal bosons in the same harmonic trap, since $\sigma_{k}^{\rm ideal}=\sqrt{m^* {\omega^*}^2/ k_{B}T}=\sqrt{m \omega^2/ k_{B}T}$. It is thus possible to use an optical lattice to enhance the strength of interactions and reach the low-temperature regime $k_{B} T \ll \mu$.

The experiment starts with the production of a $^4$He$^*$ BEC with $N=40 (4)\times 10^3$ atoms which is loaded in a 3D optical lattice of amplitude $V=9.5~E_{R}$ \cite{cayla2018}, with $E_{R}/h=h/8md^2=20.7~$kHz and $d=775$ nm. The overall harmonic trap is isotropic, with a frequency $\omega/2 \pi = 300(20)$~Hz. By tuning the lattice amplitude, we set $U/J \simeq 10$ ($U/h = 4350~$Hz and $J/h = 450~$Hz). The interaction energy is $\mu=n_{0} U$, where the lattice filling $n_{0}$ at the trap center is close to one in this work ($0.9 \leq n_{0}\leq 1.6$). The critical temperature for Bose-Einstein condensation is $T_{\rm BEC} =5.9(2)J/k_{B}$ \cite{cayla2018}.

\begin{figure}[ht!]
\includegraphics[width=\columnwidth]{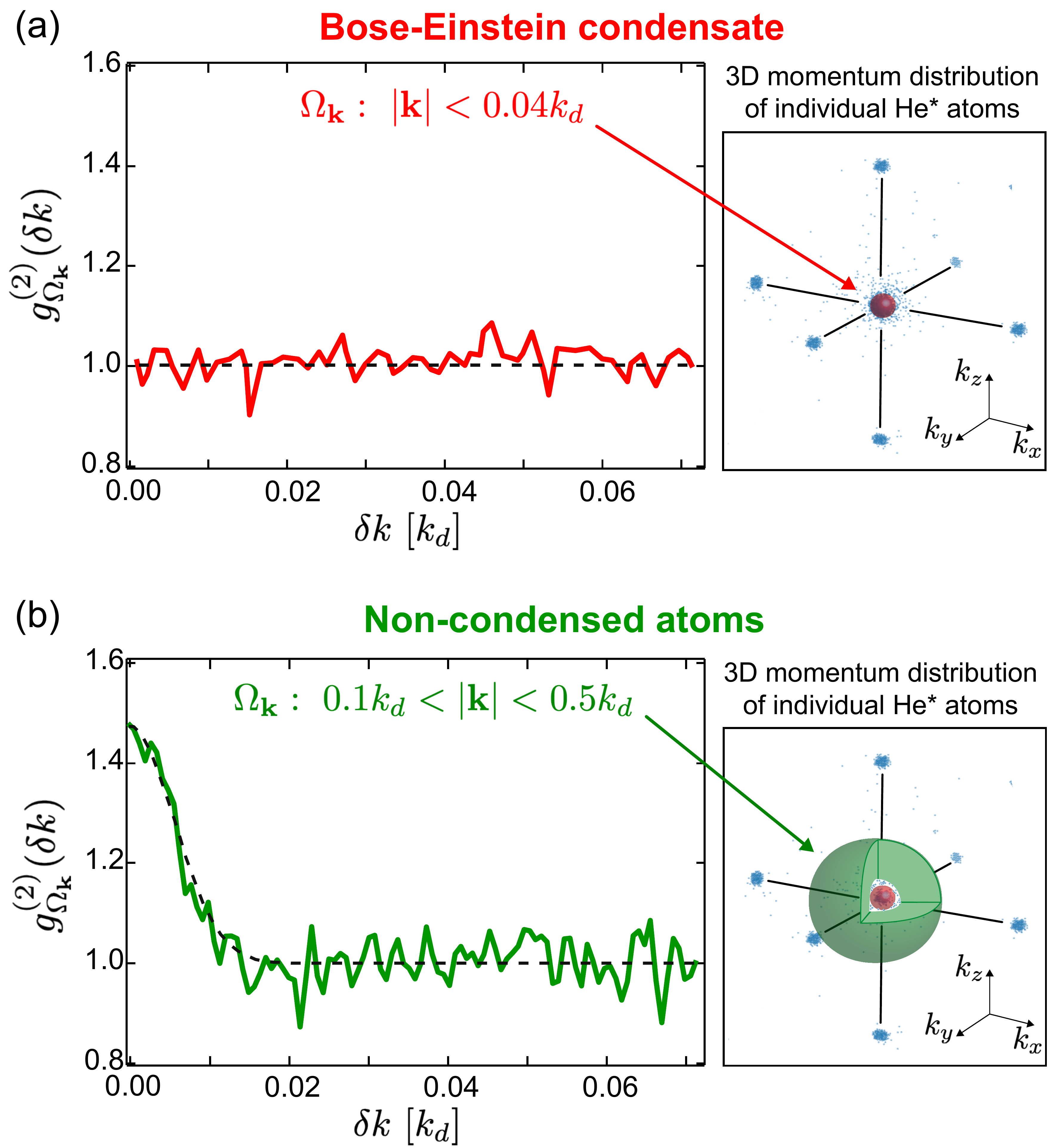}
\caption{Two-body HBT correlations in a strongly interacting lattice Bose gas at $T=2.9J$. The plots are 1D cuts through the 3D correlation function   along the lattice axis ${\bf u}_{x}$ with $\Delta k _{\perp}=10^{-2} k_{d}$. (a) Two-body correlation function $g_{\Omega_{\bf k}}^{(2)}(\delta k)$ in the condensate.  Inset: the red sphere depicts the volume $\Omega_{\bf k}$ ($|{\bf k}| <0.04 k_{d}$) over which the correlations are calculated. We find $g_{\Omega_{\bf k}}^{(2)}(\delta k_{x})=1.0(1)$ for the condensate mode, {\it i.e.} no bunching as expected when one mode only is populated.  (b)  $g_{\Omega_{\bf k}}^{(2)}(\delta k)$ in the non-condensed fraction. Inset: the green region depicts the volume $\Omega_{\bf k}$ ($0.1 k_d < | {\bf k}| <0.5 k_{d}$) over which the correlations are calculated. Note that the condensate is excluded from this volume. One observes a well contrasted bunching whose bell shape is fitted with a Gaussian function (dashed line) to quantify the bunching properties. 
}
\label{fig2}
\end{figure}

We measure 3D single-atom-resolved distributions with the He$^*$ detector after a free fall of $\sim 325$~ms (see Fig.~\ref{fig1}b) \cite{cayla2018}. Since interactions do not affect the expansion from a lattice with less than two atoms per site \cite{cayla2018, tenart2020}, the long free fall maps the in-trap momentum distribution on the measured spatial distribution. Recording the 3D momentum distributions provides a natural separation of the condensate from its depletion. Indeed, the ${\bf k}$-space density of a lattice BEC is made of periodically-spaced (period $k_{d}$) sharp peaks of width $\sim 1/R_{\rm bec}$, while the non-condensed fraction -- quantum depletion and thermal phonons -- extends over the entire Brillouin zone of width $k_{d}$. Because $R_{\rm bec}\simeq 23\, d \gg d$ here, the contribution of the non-condensed fraction is negligible in the ${\bf k}$-space volume $\sim R_{\rm bec}^{-3}$ occupied by the condensate. We exploit this property to perform the integral of Eq.~\ref{eq:g2} over different volumes $\Omega_{\bf k}$, which allows us to investigate the HBT correlations in the two components separately (see Fig.~\ref{fig2}). We determine the correlation properties along one lattice axis at a time, with a small transverse integration of $\pm \Delta k _{\perp} \leq 1/R_{\rm bec}$ to increase the signal-to-noise ratio \cite{SuppMatt}. 

An example of measured correlation functions $g_{\Omega_{\bf k}}^{(2)}(\delta k)$ in the two components is plotted in Fig.~\ref{fig2}. We find that $g_{\Omega_{\bf k}}^{(2)}(\delta k)$ is constant and equal to 1 in the condensate (see Fig.~\ref{fig2}a), {\it i.e.}, no bunching is observed. This is consistent with the fully coherent nature of the condensate \cite{aspect2016}. In contrast, a well-contrasted bunching is visible in the non-condensed fraction (see Fig.~\ref{fig2}b).

\begin{figure}[ht!]
\includegraphics[width=\columnwidth]{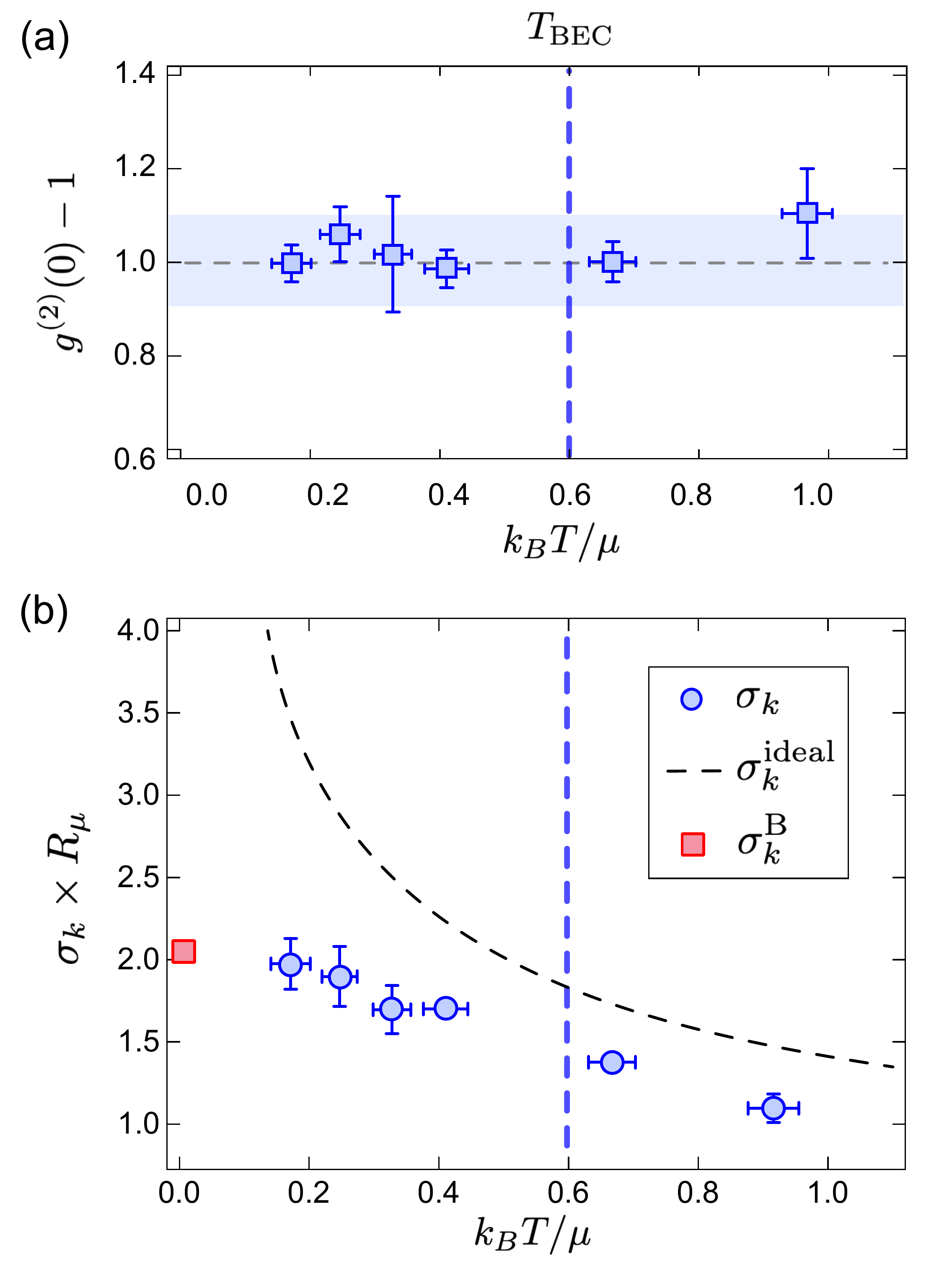}
\caption{(a) Bunching amplitude $g^{(2)}(0)-1$ as a function of the reduced temperature $k_{B}T/\mu$. The measurements are consistent with $g^{(2)}(0)= 2$, {\it i.e.} with a chaotic statistics at any temperature. The vertical blue dashed line in both panels signals $T_{\rm BEC}$. (b) Two-body correlation width $\sigma_k$ plotted as a function of $k_{B}T/\mu$. For each temperature, the value of the temperature-dependent interaction energy $\mu(T)= n_0(T) U$ in the experiment is calculated from the lattice filling $n_{0}$ at the trap center. $\sigma_k$ is expressed in units of $(R_{\mu})^{-1}=\sqrt{m \omega^2/2 \mu(T)}$. $R_{\mu}$ coincides with the BEC radius at $T=0$, $R_{\mu}(T=0)=R_{\rm bec}$. The red square corresponds to a numerical calculation of the correlation width $\sigma_k^{\rm B}$ for a harmonically-trapped {\it one-dimensional} interacting Bose gas in the Bogoliubov approximation, with parameters consistent with the {\it three-dimensional} experiment (see main text and \cite{SuppMatt}). The dashed black line is the prediction $\sigma_k^{\rm ideal}=\sqrt{\hbar \omega/k_{B}T}$ for non-interacting bosons at thermal equilibrium, for which $\sigma_k^{\rm ideal} R_{\mu} = \sqrt{2 \mu/k_{B}T} $. } 
\label{fig3}
\end{figure}

To exploit our data, we fit the bell-shaped 1D cuts $g_{\Omega_{\bf k}}^{(2)}(\delta k_j)$ ($j=\{ x, y, z\}$) of the bunching bump along the reciprocal lattice axes with Gaussian functions. We find that $g_{\Omega_{\bf k}}^{(2)}$ is isotropic, a property consistent with the isotropy of the trap geometry. In the analysis of the bunching bump, we account for the transverse integration $\Delta k_{\perp}$ and the resolution of the He$^*$ detector (rms width $\sigma=2.8(3)\times 10^{-3} k_{d}$) to extract the bunching amplitude $g^{(2)}(0)-1$ and width $\sigma_k$ \cite{SuppMatt}. Note that the data shown in Fig.~\ref{fig2}b is the raw data before the deconvolution with the point spread function of the detector, {\it i.e.} the amplitude of the bump in Fig.~\ref{fig2}b is smaller than the bunching amplitude  $g^{(2)}(0)-1$ shown in Fig.~\ref{fig3} because of the resolution of the detector. Based on this protocol, we investigate the bunching properties across the BEC transition, while keeping the ratio $U/J=10$. The temperature $T$ is varied by heating the gas in a reproducible manner \cite{cayla2018} and calibrated by comparison with ab-initio Quantum Monte-Carlo calculations \cite{SuppMatt}. The results are plotted in Fig.~\ref{fig3}. 

Firstly, we find that the bunching amplitude is constant with temperature and equal to $g^{(2)} (0) -1 = 1.0(1)$ (see Fig.~\ref{fig3}a). At large temperatures $T>T_{\rm BEC}$, this observation corresponds to the usual HBT bunching of weakly-interacting thermal bosons. Below $T_{\rm BEC}$, the bunching can be understood within Bogoliubov theory, along the lines of our introduction. Indeed, the Bogoliubov approximation is believed to be accurate at $k_{B}T\ll\mu$: in homogenous systems without a lattice \cite{giorgini1999, lopes2017}, it was shown to be reliable up to values of the quantum depletion ($\sim15\%$) similar to that of our experiment. Measuring $g^{(2)}(0)\simeq 2$ for $k_{B}T/\mu\leq0.4$ confirms the chaotic statistics of the Bogoliubov phonons. Moreover, this result extends to temperatures as low as $k_{B }T / \mu = 0.17$ where an equal fraction of atoms belong to the quantum depletion and to the Bogoliubov phonons. It suggests that $g^{(2)}(0)=2$ also for the quantum depletion, albeit for the different mechanism sketched in the introduction, {\it i.e.} a partial trace over the atom pairs in the quantum depletion. 

Secondly, the bunching width $\sigma_k$ is systematically smaller than that of ideal bosons  $\sigma_k^{\rm ideal}$, in the same trap at the same temperature (see Fig.~\ref{fig3}b). This difference is more pronounced at small values of $k_{B}T/\mu$ as a result of interactions. For $T>T_{\rm BEC}$, one expects to observe the width corresponding to a thermal gas with interactions that broaden the in-trap size with respect to that of an ideal thermal gas. This prediction is compatible with our observation of $\sigma_k$ slightly below $\sigma_k^{\rm ideal}$. Note that we could not increase the temperature beyond $k_{\rm B} T \sim 0.9 \mu$ while keeping the atoms in the lowest lattice band. In the opposite low-temperature regime, the value $\sigma_k \simeq 2/R_{\mu}$ corresponds to an in-trap size close to that of the condensate $R_{\rm bec} = R_{\mu}(T=0)$. To be quantitative, we have numerically solved the simplified case of a trapped 1D interacting Bose gas in the Bogoliubov approximation with parameters consistent with our 3D experiment. More specifically, we use in the numerics the ratio $\mu/\hbar \omega=51$ identical to that of the experiment $\mu/\hbar \omega^*=\langle n_{0} \rangle U/\hbar \omega^*$, and the 1D integral of Eq.~\ref{eq:g2} is calculated for the non-condensed atoms only (excluding the region $k R_{\rm bec}<10$) using a 3D-like weight $\propto k^2$ \cite{SuppMatt}. The numerical result $\sigma_k^{\rm B}$ (red square) is compatible with our measured low-temperature $\sigma_k$. Since the value of $\sigma_k^B$ crucially depends on the collective nature of the excitations and can be unambiguously attributed to the spatial extension of the Bogoliubov phonons and quantum depleted atoms within the condensate \cite{SuppMatt}, we ascribe the measured width to the same physical origin. A quantitative comparison with theory at any temperature would require more sophisticated techniques that go beyond the Bogoliubov approximation used here.

In conclusion, we have observed and fully characterized the atom bunching occurring in the non-condensed fraction of a strongly interacting Bose gas. We have shown that this bunching phenomenon directly reflects the interplay of interactions and quantum statistics, through the properties of phonons and of the quantum depletion. Our results thus demonstrate that momentum-momentum correlations provide information about the quantum state of strongly interacting bosons, extending the interest of  HBT-like experiments beyond the case of ideal particles. This method will be used to look for two-body correlations at opposite momenta that are expected for the quantum depletion and other many-body phenomena \cite{altman2004}. Such a measurement will demand to achieve the large signal-to-noise required at finite temperature \cite{mathey2009}, but will be of great importance to directly reveal pairing mechanisms.

\vspace{5mm}
\begin{acknowledgments}
We are grateful to Giuseppe Carleo for providing us with the QMC calculations for our experiment. We acknowledge fruitful discussions with Antoine Browaeys and the members of the Quantum Gases group at Institut d'Optique. This work benefited from financial support by the LabEx PALM (Grant number ANR-10-LABX-0039), the R\'egion Ile-de-France in the framework of the DIM SIRTEQ, the ``Fondation d'entreprise iXcore pour la Recherche" and the Agence Nationale pour la Recherche (Grant number ANR-17-CE30-0020-01). I. C. acknowledges from the European Union H2020-FETFLAG-2018-2020 project ``PhoQuS" (n.820392) and the Provincia Autonoma di Trento. S. B. acknowledges funding from the Leverhulme Trust Grant No. ECF-2019-461 and the Lord Kelvin/Adam Smith (LKAS) Leadership Fellowship. A. A. holds the Augustin Fresnel Chair of Institut d'Optique Graduate School, supported by Nokia Bell Labs.
\end{acknowledgments}

\cleardoublepage
\begin{widetext}
\begin{center}
\textbf{\large Supplemental Material: Hanbury-Brown and Twiss bunching of phonons and of the quantum depletion in a strongly-interacting Bose gas}
\\
\vspace{5mm}
H. Cayla, S. Butera, C. Carcy, A. Tenart, G. Herc\'e, A. Aspect, I. Carusotto and D. Cl\'ement
\vspace{5mm}
\end{center}
\end{widetext}

\setcounter{figure}{0} 
\setcounter{equation}{0} 
\renewcommand\theequation{S\arabic{equation}} 
\renewcommand\thefigure{S\arabic{figure}}  

{\bf Experimental values of the temperature and the quantum depletion.} We have performed quantum Monte-Carlo (QMC) calculations at finite temperature and with the experimental parameters, including the trap inhomogeneity, as described in \cite{cayla2018}. Note that our implementation of the QMC calculations yield the in-trap and momentum densities but not the two-body correlations such as those measured in the experiment. The temperature $T$ of the lattice gases is the only parameter that is not directly measured in the the experiment, and it is thus extracted from fitting the measured momentum distributions with those calculated at various temperatures by means of QMC \cite{cayla2018}. The values of the lattice filling $n_{0}$ at the trap center ($0.9 \leq n_{0} \leq 1.6$ in the experiment) and of the quantum depletion ($\sim 15\%$ at $k_{B}T\mu \ll 1$) are obtained from the numerical results of the QMC calculations as well. 
\\

{\bf Calculation of the volume integrated two-body correlations $g_{\Omega_{\bf k}}^{(2)}(\delta {\bf k})$.} From the 3D distributions of individual atoms measured with the He$^*$ detector, we extract the two-body correlations over a ${\bf k}$-space volume ${\Omega_{\bf k}}$ proceeding with the approach described in \cite{carcy2019}. It consists in calculating the histogram of the distances $\delta k$ of one atom belonging to ${\Omega_{\bf k}}$ with all the other atoms from the same run of the experiment in the volume ${\Omega_{\bf k}}$, and then summing all the histograms belonging to the same run. This approach is performed along a specific axis ${\bf u}$ of the momentum space where the distance between two atoms is evaluated, $\delta k=\delta {\bf k}. {\bf u}$, with $\delta {\bf k}={\bf k}_{1}-{\bf k}_{2}$ where ${\bf k}_{1}$ ({\it resp.} ${\bf k}_{2}$) is the momentum of the first ({\it resp.} second) atom. In addition, we calculate the histograms of the distance with a transverse integration $\Delta k_{\perp}$ in order to raise the signal-to-noise.
\\

\begin{figure}[ht!]
\includegraphics[width=\columnwidth]{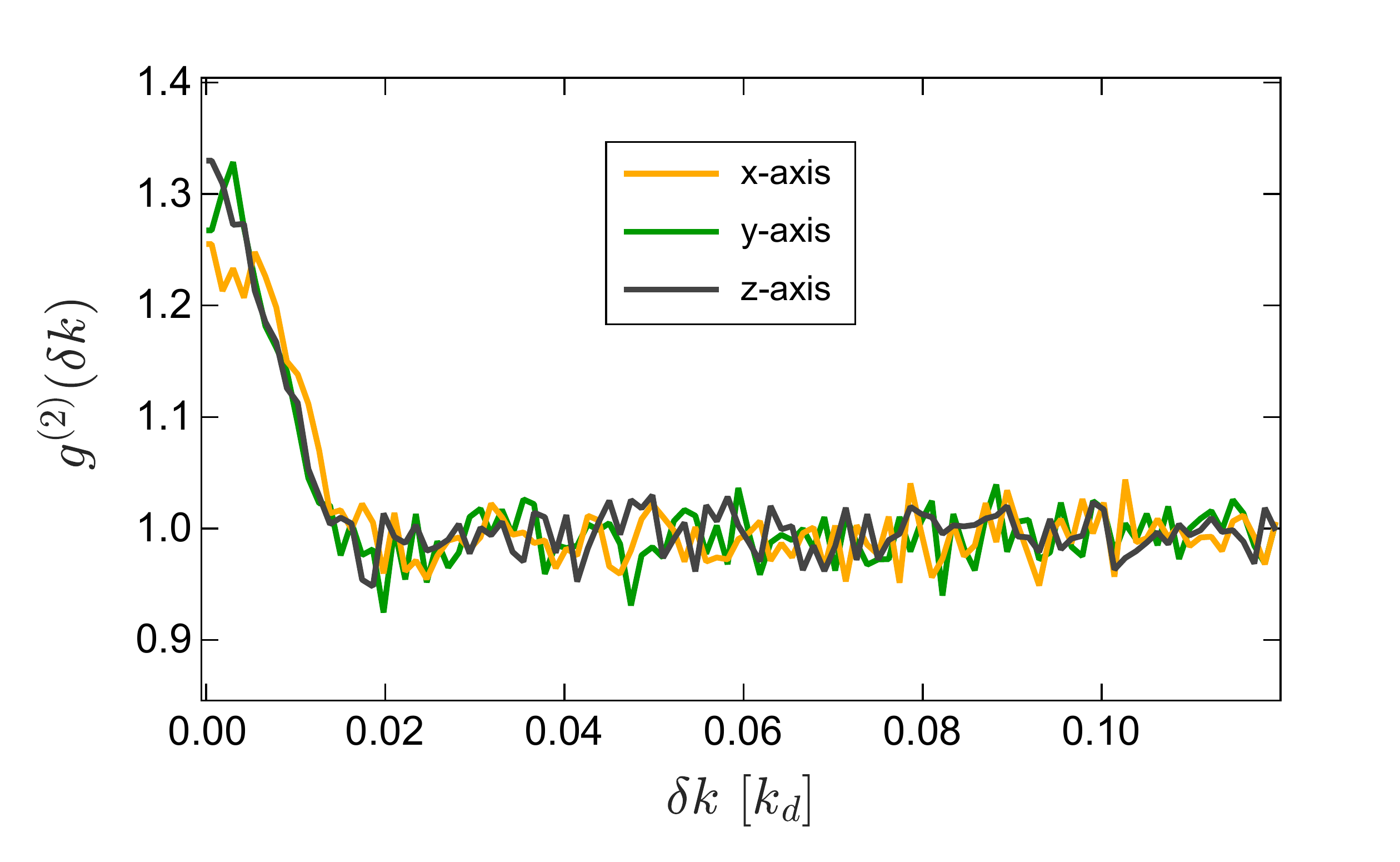}
\caption{Plot of $g^{(2)}(\delta k)$ along the three axis of the reciprocal lattice. This data set was recorded at $T=8.7J$ and it is plotted for a transverse integration $\Delta k_{\perp}= 1.2 \times10^{-2} k_{d}$. This comparison illustrates that the measured correlation functions are well fitted by an isotropic Gaussian function.} 
\label{fig1-sup}
\end{figure}

{\bf Isotropy of the measured correlations $g^{(2)}(\delta {\bf k})$.} The correlation signals presented in this work have been measured along the eigenaxis of the reciprocal lattice and the transverse integration used is varied from $\Delta k_{\perp}=5 \times 10^{-3} k_{d}$ to $\Delta k_{\perp}=5 \times 10^{-2} k_{d}$. The histograms of the distances $\delta k$ is calculated with a fixed resolution $\Delta k=1.2 \times 10^{-3} k_{d}$. In Fig.~\ref{fig1-sup}a, we plot the two-body correlation functions $g^{(2)}(\delta k)$ along the three eigenaxis of the lattice from one set of data ($T=8.7J$). This example illustrates that the measured correlations are isotropic, a property that we use in the quantitative analysis of the HBT peaks.
\\

\begin{figure}[h!]
\includegraphics[width=\columnwidth]{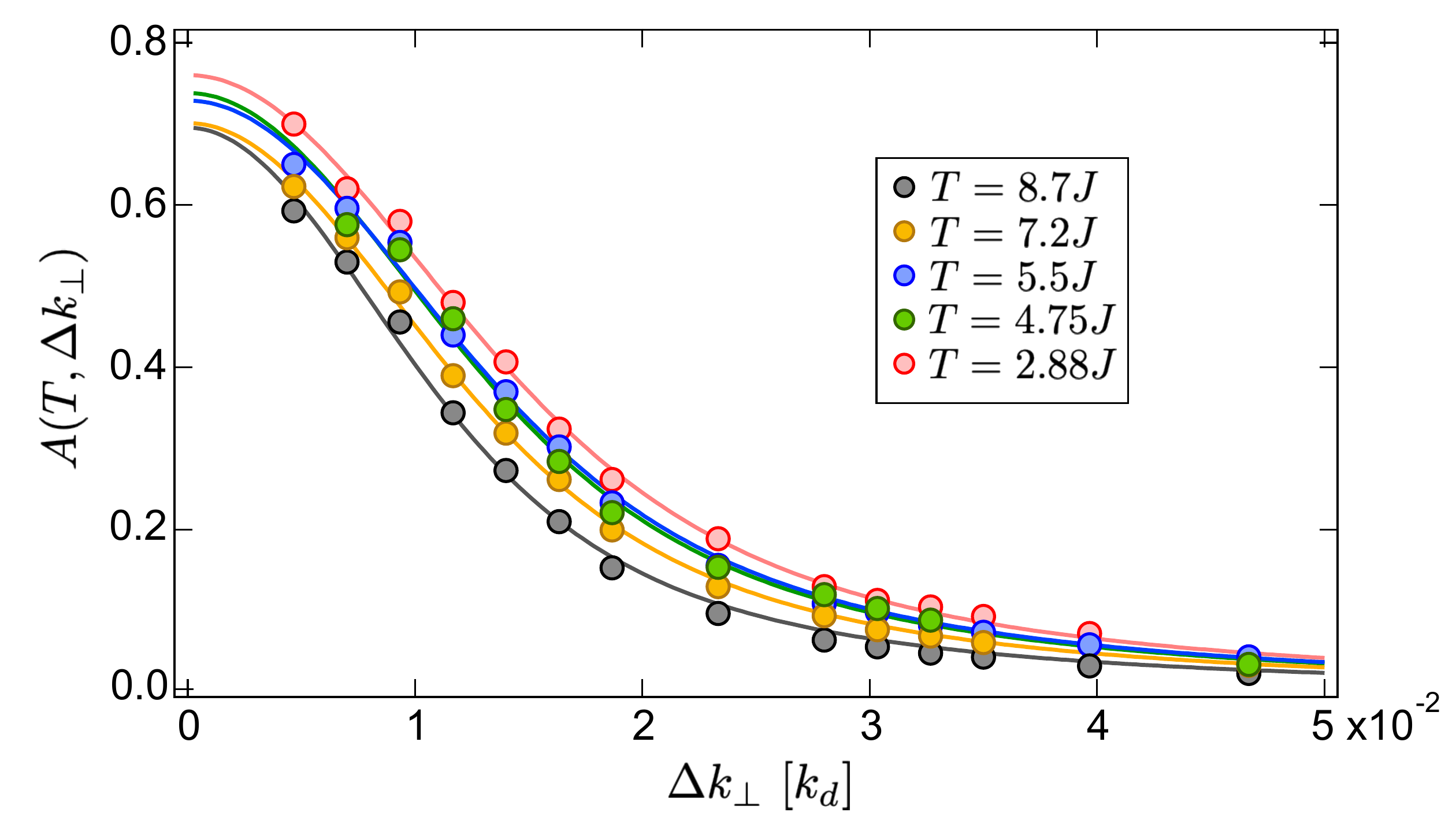}
\caption{Plot of the measured amplitude $A(T,\Delta k_{\perp})$ of the measured bunching bump as a function of the transverse integration $\Delta k_{\perp}$ for different temperatures $T$. The solid lines are a fit of the amplitude $A$ as a function of the transverse integration $\Delta k_{\perp}$ that yields the amplitude $A_{0}$, expected for a vanishing transverse integration, and the correlation length (see text).} 
\label{fig2-sup}
\end{figure}

{\bf Analysis of the bunching amplitude and width.} For each value $\Delta k_{\perp}$ of the transverse integration (in the range 0.5 -- 5$\times10^{-2}k_{d}$), we extract the amplitude $A$ and the width $\Sigma_k$ from fitting $g_{m}^{(2)}(\delta k)$ with $G(\delta k) = 1 + A \, {\rm exp}\left(-\delta k^2/ \Sigma_k^2 \right)$. We have chosen a small longitudinal discretisation $\Delta k= 1.2 \times 10^{-3} k_{d}$ which is much smaller that the fitted values of $\Sigma_k$ to ensure that $\Delta k$ does not affect the fitting of $\Sigma_k$. In addition, we keep $\Delta k$ fixed when varying the transverse integration $\Delta k_{\perp}$. In Fig.~\ref{fig2-sup} we plot the amplitude $A$ as a function of $\Delta k_{\perp}$ for the various temperatures, along with the fitting function of Eq.~\ref{eq:fit-h}.

We find that $\Sigma_k$ does not depend on $\Delta k_{\perp}$, from which we conclude that the correlations are separable. On the contrary, the amplitude of the bunching peak depends on the transverse integration $\Delta k_{\perp}$ and we fit this dependency with that expected for a 3D isotropic function,
 \begin{equation}
 \label{eq:fit-h}
A(\Delta k _{\perp}) =  A_{0} \times \frac{\pi \Sigma_k^2}{(2 \Delta k_{\perp})^2} \left [ {\rm erf}\left (\frac{ \Delta k_{\perp}}{\Sigma_k} \right ) \right]^2 
\end{equation}
where ${\rm erf}(x) = \frac{2}{\sqrt \pi}\int_0^x {\rm exp}(-y^2)dy$ is the error function. This yields the value of $A_{0}(T)$ of the bunching amplitude at vanishingly small transverse integration $\Delta k_{\perp}$.
\\

{\bf Effect of the finite resolution of the detector.} The finite resolution of the detector affects the measured values of the amplitude $A_{0}$ and of the correlation length $\Sigma_k$ when the latter is not much larger than the resolution. Assuming that the (one-particle) point spread function of the detector is a Gaussian function with a RMS width $\sigma$, the measured two-body correlation length $\Sigma_k$ writes $\Sigma_k= \sqrt{\sigma_k^2+4\sigma^2}$ where $\sigma_k$ is the two-body correlation length of the gas. The effect of the resolution on the measured amplitudes of the bunching $A_{0}$ corresponds to an integration over a volume $\propto \sigma^3$.

The calibration of the detector (MCP from the Burle company, channel diameter 25$\mu$m and angle $8^\circ$) resolution was performed with a thermal gas of atoms in a harmonic trap.. For the experimental parameters we use, the bunching amplitude  varies between $g^{(2)}(0)=1.05$ and $g^{(2)}(0)=1.45$ as a function of the transverse integration. The two-body correlation lengths are different along the different trap axis because of the anisotropy in the trapping frequencies. The expected values $\ell_{c,j}=1/s_{j}=\sqrt{m \omega_{j}^2/k_{B}T}$ for a non-interacting gas in the optical trap we use are $\ell_{c,x}=29~$mm$^{-1}$,  $\ell_{c,y}=103~$mm$^{-1}$ and  $\ell_{c,z}=89~$mm$^{-1}$. The measured correlation lengths $\ell_{c,x}\simeq 33(8)~$mm$^{-1}$, $\ell_{c,y}\simeq 108(16)~$mm$^{-1}$ and $\ell_{c,z}\simeq 91(10)~$mm$^{-1}$ reproduce the trap anisotropy and are compatible with the expected ones. 

Finally, we extract the bunching amplitude $g^{(2)}(0)-1=A_{0} \times (\Sigma_k/\sigma_k)^3$ and width $\sigma_k= \sqrt{\Sigma_k^2-4\sigma^2}$ of the interacting gas accounting for the resolution of He$^*$ detector.
\\

\begin{figure}[ht!]
\includegraphics[width=\columnwidth]{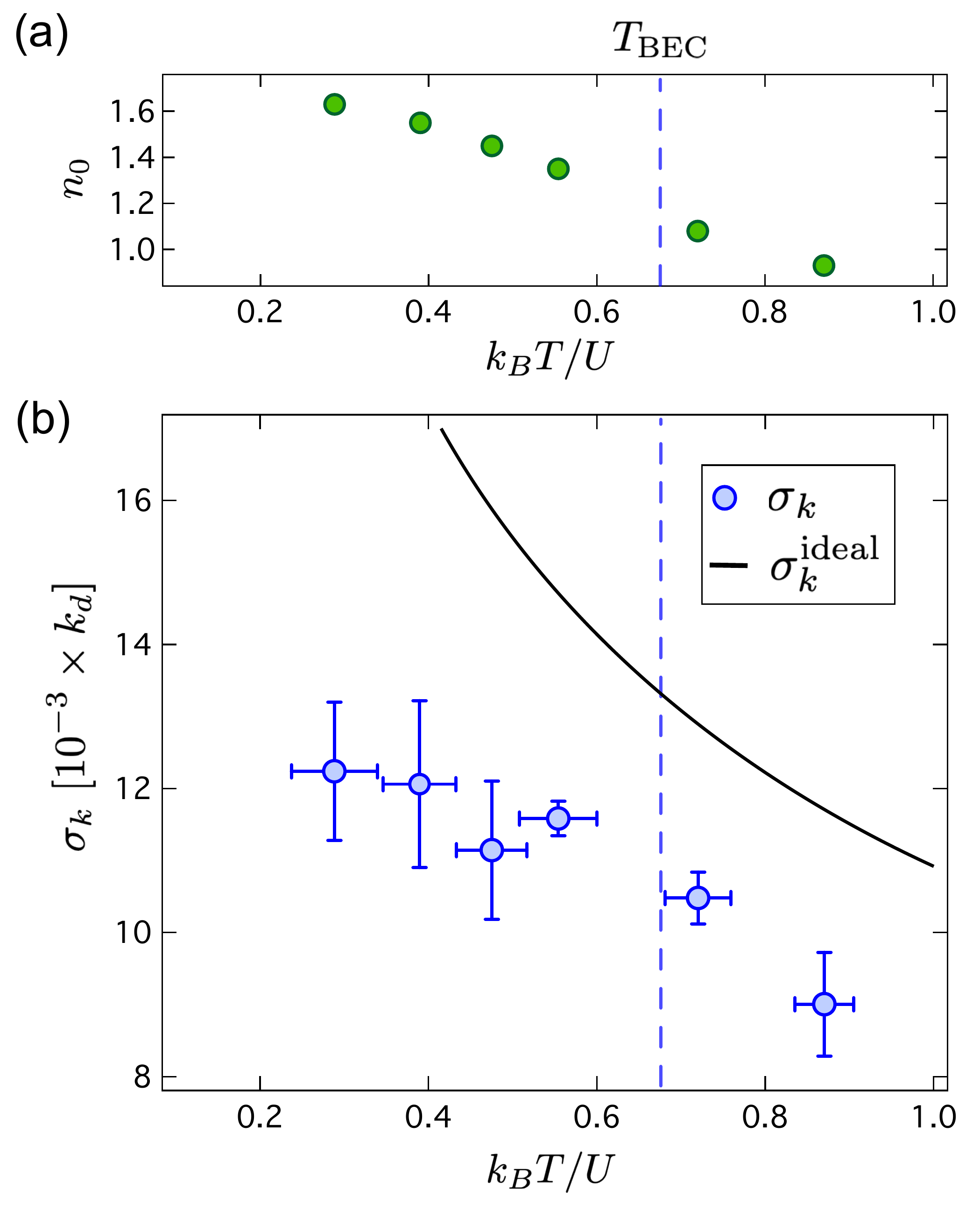}
\caption{{\bf (a)}, Lattice filling $n_{0}$ at the trap center obtained from the QMC calculations with the experimental parameters. {\bf (b)}, Bunching width $\sigma_k$ as a function of the reduced temperature $k_{B}T/U$ (green dots). The solid black line corresponds to the prediction $\sigma_k^{\rm ideal}$ for non-interacting bosons at thermal equilibrium, rescaled by the interaction energy $U$.} 
\label{fig3-sup}
\end{figure}

{\bf Bunching width $\sigma_{k}$.} Here we plot  $\sigma_k$ as a function of the on-site interaction energy $U$ in Fig.~\ref{fig3-sup}b as an alternative, along with the lattice filling $n_{0}$ at the trap center for the various experimental parameters in Fig.~\ref{fig3-sup}a. Note the lattice filling $n_{0}$ is decreasing with increasing temperatures because we keep the total number of atoms constant. 

The prediction $\sigma_k^{\rm ideal}$ for an ideal Bose gas in a harmonic trap is independent of the presence of the lattice. Indeed, the effective mass $m^*$ and frequency $\omega^*$ are such that $m \omega^2 =m^* {\omega^*}^2$. On the contrary, the numerical calculations for trapped 1D interacting bosons depend on the ratio $\mu / \hbar \omega$, the value of which depends on the presence, or absence, of the optical lattice. Therefore, we have matched this ratio $\mu / \hbar \omega$ for the numerics, performed in the absence of a lattice, with the ratio $\mu/ \hbar \omega^*=\langle n_{0}\rangle U / \hbar \omega^*$ of the experimental configuration.
\\

{\bf Numerical Bogoliubov calculations.} We evaluated the zero temperature limit of the bunching width $\sigma_k$, reported in Fig.~\ref{fig3}, by numerically solving the Bogoliubov theory in the one-dimensional case. The Bogoliubov theory is the first-order correction of the many-body problem beyond the mean-field Gross-Pitaevskii (GP) description and is valid for weakly-interacting systems. The harmonic trap potential is explicitly included in our calculation: first we evaluate the GP ground state for the condensate at the chemical potential of the experiment, that is $\mu/\hbar\omega=\langle n_{0} \rangle U/\hbar \omega^*=51$. Then, we numerically obtain the eigenmodes of the linearized Bogoliubov-de Gennes equations on top of the trapped condensate. The momentum-space correlation function is finally evaluated by decomposing the four-operator averages of the non-condensed atoms into products of two-operator ones by means of the Wick theorem and then calculating these latter on a low-temperature thermal (thus Gaussian) state. To reproduce the 3D case of the experiment, we integrated the 1D analog of Eq.~(1) of the main text with a $k^2$ weight in the 1D integrals. Similarly to the experiment, we removed the $k$-space region of radius $k R_{\rm bec}<10$ around the $k=0$ position of the condensate, in order to restrict ourselves to the non-condensed component only.
\\

\begin{figure}[ht!]
\includegraphics[width=\columnwidth]{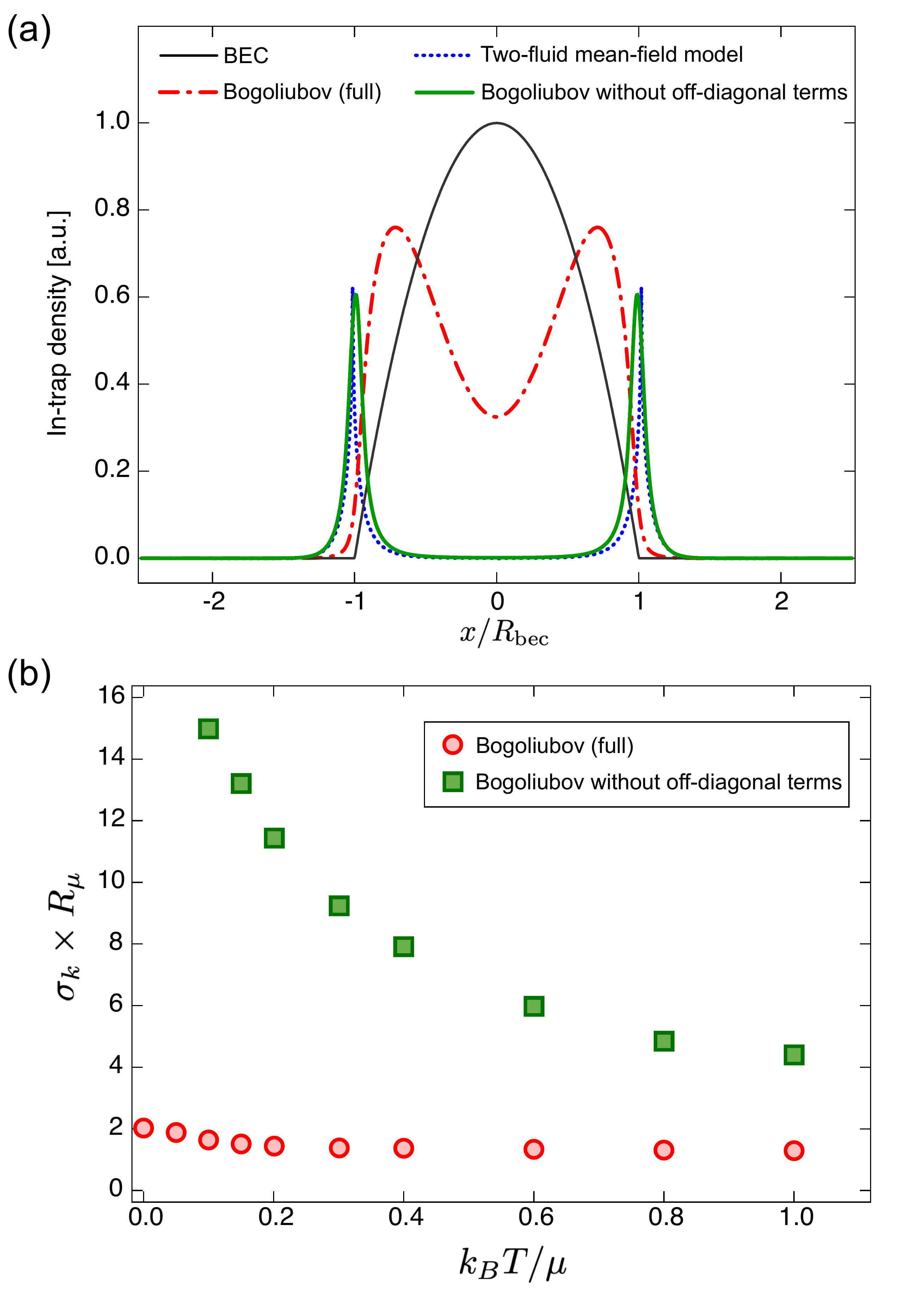}
\caption{{\bf (a)}, In-trap densities of the condensate (black solid line) and of the non-condensed component calculated in various approaches: the (full) Bogoliubov theory (red dot-dashed line), the single-particle approximation to the Bogoliubov theory, that results by turning-off the off-diagonal couplings in the Bogoliubov-de Gennes equations (green solid line), and the two-fluids mean-field model (dotted blue line). For readability sake, we have rescaled the density of the (full) Bogoliubov theory by a factor 1/10 with respect to the densities of the non-condensed components in the other two approaches. The density profile of the full Bogoliubov approach reflects the collective character of the low energy Bogoliubov modes, which are delocalized inside the condensate, and the presence of the quantum depletion. This contrasts with the other two density profiles associated to single-particle excitations, that are localized at the trap edge.  {\bf (b)}, Numerical solution of the bunching width $\sigma_{k}^{\rm B}$ in the Bogoliubov approximation (red dots) and in the Bogoliubov approximation without the off-diagonal terms (green square).} 
\label{fig4-sup}
\end{figure}

{\bf Origin of the bunching: collective vs single-particle thermal excitations.} 
In this work we have used the concept of quasi-particles in the Bogoliubov approximation to treat the effect of interactions and argue that our observations are compatible with the two-body correlations expected in the presence of collective excitations -- Bogoliubov phonons -- and the quantum depletion. To better illustrate this origin of the observed atom-atom correlations, we now contrast the measured bunching width $\sigma_{k}$ with that expected for single-particle thermal excitations, in the presence of interactions. To this aim, we consider two approaches: {\it (i)} a mean-field description of two interacting fluids, the condensate and a thermal gas of single-particle excitations; {\it (ii)} a further approximated version of the Bogoliubov theory where off-diagonal terms are neglected, so that the Bogoliubov modes lose their collective character and reduce to single-particle excitations.

We first consider a two-fluid model that describe the condensate and a thermal gas of single-particle excitations. The interactions between the thermal gas and the condensate are treated at the mean-field level. In this approach, the repulsive potential created by the condensate compresses the thermal gas to a region of size much smaller than $R_{\rm bec}$ at the edge of the trap (see Fig.~\ref{fig4-sup}a). The in-trap density of the thermal component writes
\begin{equation}
n_{\rm th}({\bf r})\propto\frac{1}{\lambda_{dB}^3} \ g_{3/2} \left (- \frac{\mu}{k_{B}T} \left |1-({\bf r}/R_{\rm bec})^2 \right | \right ).
\end{equation}
For a ratio $k_{B}T/\mu \simeq 0.17$ as probed in the experiment, the half-width half-maximum of the in-trap density for the thermal gas equals $\simeq R_{\rm bec}/20$. The HBT bunching associated with this compressed thermal gas would have a width much larger than $1/R_{\rm bec}$ and is not compatible with our observations. 

As a further evidence in this direction, in Fig~\ref{fig4-sup}b we compare the predictions of the full Bogoliubov approximation for the temperature-dependence of $\sigma_k^B$ with the ones of a further approximated theory where we neglect the off-diagonal coupling terms in the Bogoliubov-de Gennes equations. This approach puts the two-fluid model described above on rigorous grounds by including the quantized single-particle levels in the combined potential including the trap and the mean-field repulsion by the condensate (see Fig.~\ref{fig4-sup}a). The dramatic deviation of the green squares in Fig~\ref{fig4-sup}b from the red circles of the full Bogoliubov theory (that, as shown in the main text, well captures the experiment at very low $T$) confirms the key role played by the collective effects in the experimental observations. 

As we have stressed in the main text, a direct comparison of the experimental measurements with the Bogoliubov calculation is expected to be valid only close at small temperatures $k_{B}T\ll \mu$ where the condensed fraction in the experiment is large. Nevertheless, the results plotted in Fig~\ref{fig4-sup}b clearly confirm our claims on the crucial importance of the collective (rather than single-particle) nature of the Bogliubov excitations and of the consequent quantum depletion.

\end{document}